\documentclass[preprint,12pt,numbers,sort,compress]{elsarticle}

\usepackage{amssymb}
\usepackage{amsmath}
\journal{Spectrochimica Acta, Part A: Molecular and Biomolecular Spectroscopy}

\begin{document}

\begin{frontmatter}



\title{Deciphering hidden layers's images through terahertz spectral fingerprints}


\author[1]{Candida Moffa\corref{cor1}}
\author[1]{Daniele Francescone}
\author[1]{Alessandro Curcio}
\author[1,2]{Anna Candida Felici}
\author[3]{Marco Bellaveglia}
\author[3]{Luca Piersanti}
\author[1]{Mauro Migliorati}
\author[1,4]{Massimo Petrarca\corref{cor1}}

\cortext[cor1]{Corresponding authors: candida.moffa@uniroma1.it; massimo.petrarca@uniroma1.it}

\affiliation[1]{organization={Department of Basic and Applied Sciences for Engineering (SBAI), Sapienza, University of Rome},
            addressline={Via Antonio Scarpa, 16}, 
            city={Rome},
            postcode={00161}, 
            country={Italy}}

\affiliation[2]{organization={Research Center for Applied Sciences to the Safeguard of Environment and Cultural Heritage (CIABC), Sapienza University of Rome},
            addressline={Piazzale Aldo Moro, 5}, 
            city={Rome},
            postcode={00185}, 
            country={Italy}}

\affiliation[3]{organization={INFN-LNF},
            addressline={Via Enrico Fermi, 54}, 
            city={Frascati (RM)},
            postcode={00044}, 
            country={Italy}}

\affiliation[4]{organization={Roma1-INFN},
            addressline={Piazzale Aldo Moro, 2}, 
            city={Rome},
            postcode={00185}, 
            country={Italy}}

\begin{abstract}
Terahertz (THz) radiation enables non-destructive, depth-resolved analysis of layered artworks. This study demonstrates THz multispectral imaging's ability to reveal concealed text beneath mock-up of pictorial layers, reconstructing hidden narratives at varying depths through frequency-domain analysis. Simultaneously, it maps pigment composition, providing valuable chemical information. Showcasing the power to penetrate and decipher stratified materials, this work establishes THz multispectral imaging as a crucial tool for unlocking hidden secrets and characterizing materials in Cultural Heritage artifacts.
\end{abstract}

\begin{graphicalabstract}
\includegraphics[width=\linewidth]{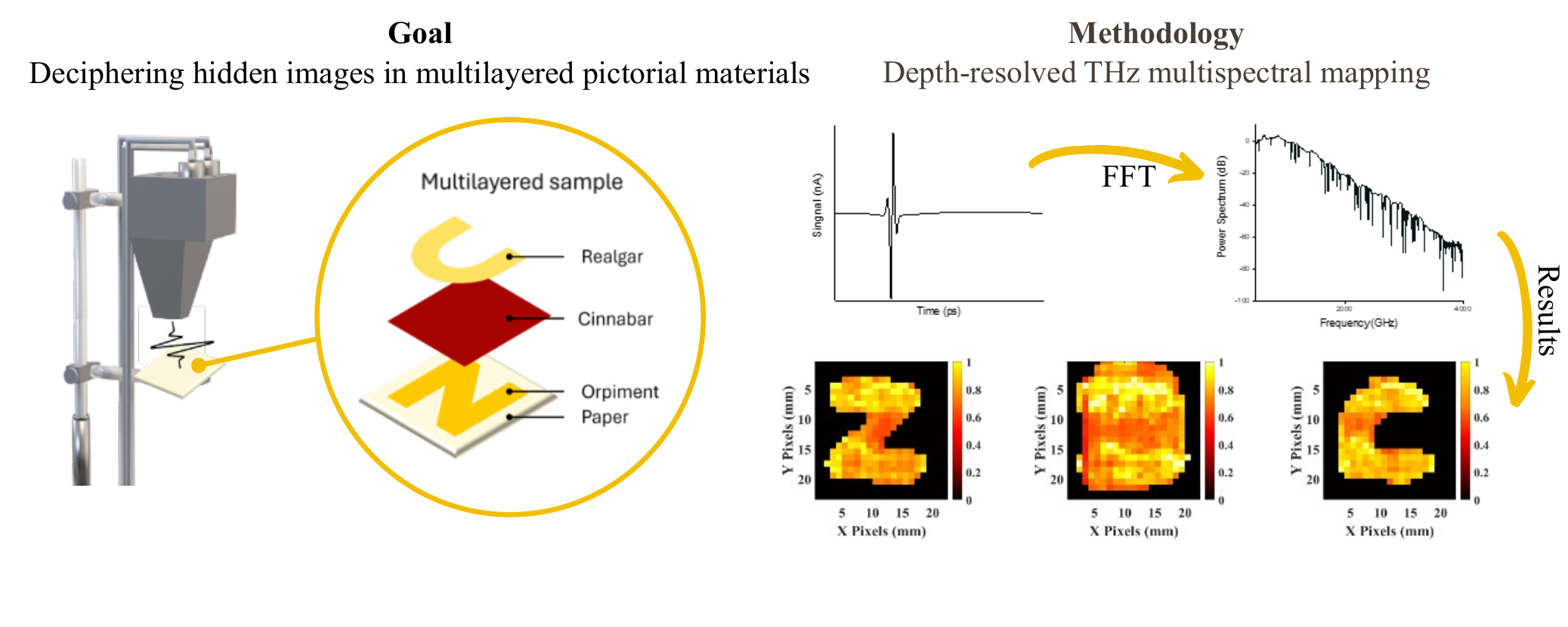}
\end{graphicalabstract}

\begin{highlights}
\item The THz-based methodology enables the simultaneous reconstruction of hidden layer images within complex layered pictorial materials, precise thickness estimation, and material differentiation using a custom spectral analysis algorithm and apparatus configuration.
\item Spectral database at THz frequencies of Cultural Heritage pigments accurately identify unique spectral fingerprints critical for the material mapping.
\item The proposed method, validated by optical microscopy and sparse deconvolution analysis, allows simultaneous reconstruction of hidden layers and precise thickness estimation.

\end{highlights}


\end{frontmatter}



\section{Introduction}
\label{sec1}
Scientific techniques have been exploited in archaeometry to investigate Cultural Heritage and are often used to unveil in a non-invasive way subsurfaces or overpainted pictorial layers.
However, chemically specific analysis of multilayers by non-destructive and non-invasive methods remains a topical issue.
Moreover, non-invasive imaging through thin stratified layers represents an analytical challenge encountered in a wide range of research potentially including polymers \cite{Frosch2010}, and biological samples \cite{Zharov2004, Conti2015b}. 

While everal methodologies, including multispectral imaging combined with X-ray fluorescence, have shown promise in recovering different layers \cite{Tonazzini2019, Perino2024}, each approach faces specific limitations related to the electromagnetic radiation employed. For instance, X-ray imaging requires sufficient elemental contrast between the layers being analyzed. Near-Infrared (NIR) spectroscopy, despite its chemical specificity based on overtone and combination bands, struggles with spectral overlap, weak absorption of certain pigments, limited penetration depth, and interference from varnishes and degradation, all of which hinder the detection of hidden layers in stratigraphic analysis. Moreover, Raman imaging can be significantly hampered by the strong fluorescence emitted by organic materials.
Recently, spatially offset Raman spectroscopy (SORS) \cite{Conti2015, Botteon2017, Vermeulen2025} has demonstrated that it is possible to recover hidden painted layers.
However, sublayer detection can be limited by intense fluorescent or pronounced Raman scattering \cite{Botteon2017} and for a multilayer system with n layers (where n$>$2), one requires n spectra obtained at different spatial offsets to retrieve individual layers \cite{Conti2014, Conti2017}.

In the context of archaeometry investigations, terahertz-based spectroscopy represents an ideal analytical technique in conservation science to study polychrome artworks; in fact, mineral pigments and organic dyes can present specific absorption vibrational modes in this spectral region \cite{Yang2017, Kleist2019, Kleist2019_2, Moffa2024, Moffa2024_2, Moffa2024_3, Moffa2025}. 
Moreover, varnishes and binders usually do not show specific fingerprints at these frequencies \cite{Fukunaga2009}; thus, it is possible to obtain selective information on the colouring materials used neglecting their contribution. 
The technique known as terahertz time-domain imaging (THz-TDI) has been exploited in several research fields \cite{Mittleman1997, Fitzgerald2005, Dandolo2017, Stubling2019} thanks to the low non-ionizing  photon energy (0.4–40 meV) that cannot cause any damage to the materials under investigation and the ability to penetrate through several stratigraphic layers. In the Cultural Heritage field, THz techniques have been used in reconstruction stratigraphy, identifying internal defects, gilded decorations, and subsurfaces \cite{Duling2009, Fukunaga2010, KochDandolo2013, Skryl2014, KochDandolo2015, FukunagaPicollo2015, Guillet2017, Catapano2017, Cacciari2018, Catapano2019, Fukunaga2023} and recovering hidden layers, drawings, or precedent pictorial cycles \cite{Adam2009, Jackson2011, Seco2013, KochDandolo2015, Gomez2017, Perino2024}. These analyses have been performed mainly using information from the temporal waveforms \cite{Lee2022}.

This study presents a THz reflection spectroscopy methodology for frequency-domain chemical mapping \cite{Shen2008, Charron2013, Bandyopadhyay2022}, specifically designed to reveal concealed images within layered pictorial materials. We initiate the process by expanding the spectral database of Cultural Heritage pigments, identifying their unique spectral fingerprints through transmission geometry spectroscopy. Subsequently, applying multispectral mapping in reflection mode to mock-up samples, we successfully reconstruct hidden images directly from frequency spectral analysis, employing a custom-developed algorithm. This method facilitates detailed stratigraphic analysis, enabling the simultaneous recovery of concealed layers, precise thickness determination, and selective material identification. The stratigraphic analysis is validated through both optical microscopy of the samples and sparse-deconvolution numerical analysis. Results from the custom-developed algorithm demonstrate a pixel recognition discrepancy of about $13\%$, which can be further refined through visual inspection of the output data. Our approach demonstrates a significantly enhanced efficacy of multispectral mapping in revealing previously unseen images within stratified artworks compared to standard temporal analysis, marking a substantial advancement in non-invasive Cultural Heritage investigations.

\section{Results}
\subsection{Pure pigments characterization}

\begin{figure}[h!]
   \centering
  \includegraphics[width=\linewidth]{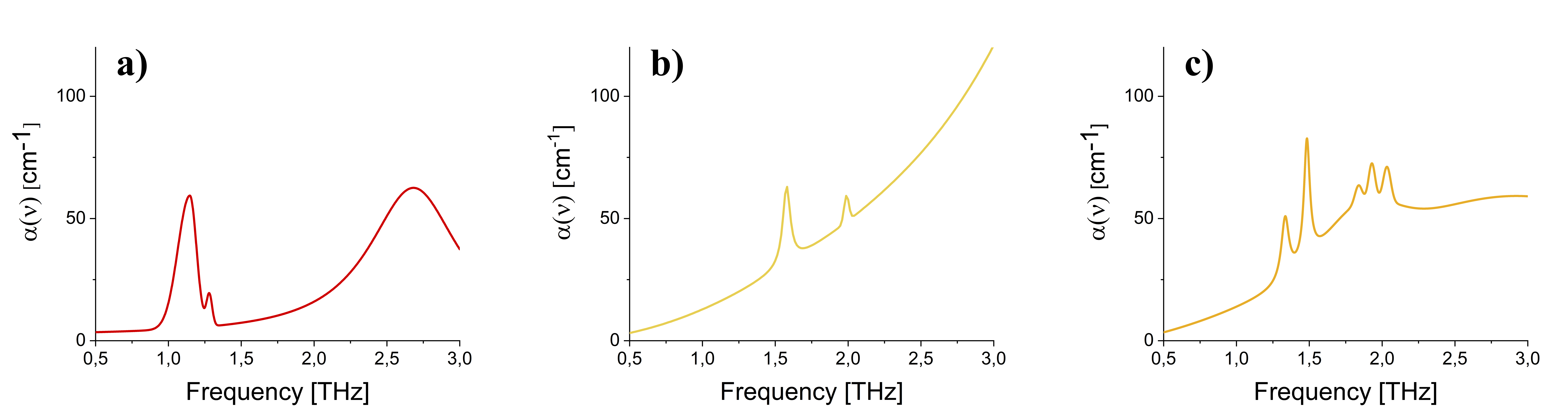}
   \caption{Lorentzian fitting of the absorbance spectrum of pure pigments a) cinnabar, b) orpiment and c) realgar in the spectral range 0.5-3 THz.}
    \label{pellet}
\end{figure}

The materials used in the preparation of the mock-up were initially characterized through terahertz time-domain spectroscopy (THz-TDS) in pressed pellets of pure pigment powder.
The absorption spectrum of cinnabar (\(HgS)\), reported in Figure \ref{pellet}a, presents specific absorption peaks at 1.17 THz, in agreement with previous research works \cite{FukunagaPicollo2015, Yang2017, Kleist2019_2, Artesani2023} This absorption fingerprint in  (\(HgS)\) has been theoretically investigated and is associated with the translation of Hg and S along the three axes of the trigonal crystal \cite{Kleist2019_2}.
The other peaks identified, centered at 1.29, and 2.66 THz were assigned to vibrations of type E \cite{Kleist2019_2}.
The refractive index of cinnabar was determined through a detailed calculation based on the observed phase shifts as it passed through the material, yielding an average value of 1.87 in the THz range.

THz spectrum of orpiment ($As_2S_3$) contains two distinct absorption peaks centered at 1.59 and 1.98 THz depending on the chain-like connection of atoms in orpiment (Figure \ref{pellet}b). The theoretical explanation for these modes has been reported \cite{Zhang2022}: the lower frequency fingerprint arises from oscillations of S and As atoms parallel with the crystallographic $bc$ plane; while the vibrational mode at 1.98 THz has been connected mainly to the perpendicular vibration with respect to the crystallographic $bc$ plane, with a strong increase in the contribution of As atoms \cite{Zhang2022}.
At THz frequencies, the retrieved refractive index for orpiment has an average value of 2.44.

Realgar ($As_4S_4$) THz absorption spectrum (Figure \ref{pellet}c) is characterized by two distinct peaks at 1.33 THz and 1.48 THz, and 1.83 THz. More specifically, the first two vibrational modes are mainly linked to the rotation of the unit cell along the crystallographic axis, while the contribution of diverse modes including the translational along the $b-$axis are responsible for the 1.83 THz vibrational mode \cite{Zhang2022}. 
To the best of the authors' knowledge, we have identified two new peaks centered at 1.92 and 2.09 THz, for which the theoretical data have been reported \cite{Zhang2022}. Expanding the database of known peaks in pigment samples is crucial, as it allows for a deeper understanding of the complex interactions between the chemical composition, molecular structure, and optical properties of the materials. 
Realgar presents a average refractive index value of 1.92 in the THz range of interest.

\subsection{Mock-ups mapping}

We prepared three mock-ups on paper (S1, S2, and S3) representing different letters with different pigments. Sample S1 shows the letter "T" in cinnabar, surrounded by a realgar background  (Figure \ref{T}a), while Sample S2 showcases the letter "H" in realgar, framed by orpiment (Figure \ref{H}a). Both pigments of S2 present a similar chemical compostion and belong to space group P21/c and the corresponding space lattice is monoclinic but they differ in the unit cell crystallizations \cite{Zhang2022}. Sample S3 consists of three superimposed layers: the base layer, composed of orpiment, forms the letter "Z" (Figure \ref{z_freq}a) which is fully obscured by a layer of cinnabar to prevent its visibility (Figure \ref{z_freq}c). On top of the cinnabar layer, the letter "C" is depicted using realgar (Figure \ref{z_freq}e).
We spatially mapped each samples in raster scan mode (see Section \ref{methods}) building an $X-Y$ matrix of the THz signals (time and frequency data)  with a resolution of \(\Delta x\)=\(\Delta y\)=1 mm corresponding to the matrix pixel dimension.

\begin{figure}[h!]
    \centering
    \includegraphics[width=0.8\linewidth]{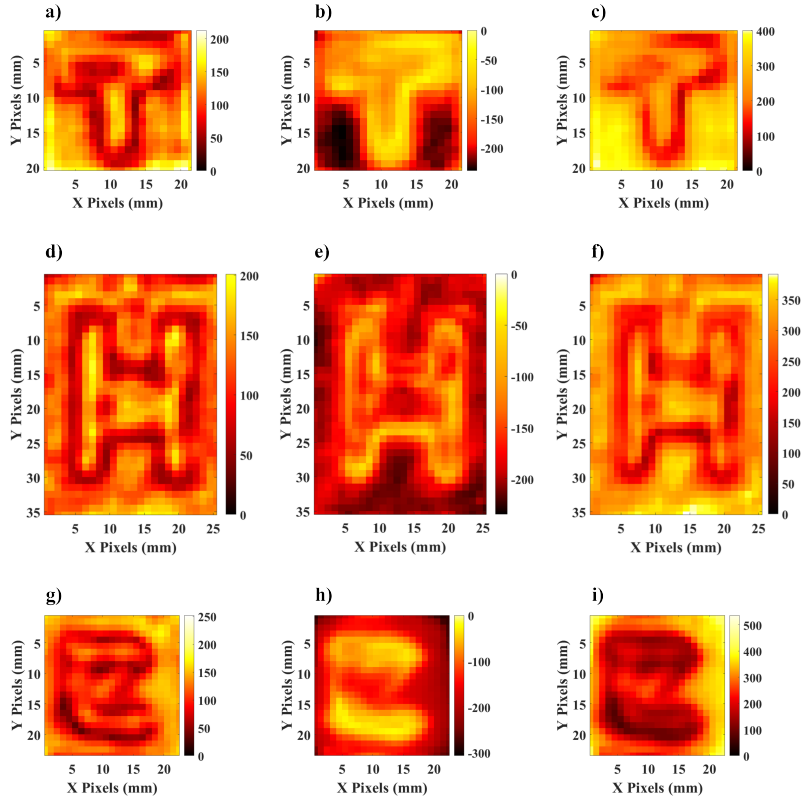}
    \caption{Image obtained from the maximum of the THz temporal waveforms for a) S1, b) S2, and c) S3; image obtained from the minimum of the THz temporal waveforms for d) S1, e) S2, and f) S3; image obtained from the peak-to-peak of the THz temporal waveforms for g) S1, h) S2, and i) S3. For all the images the pixels dimensions correspond to $\Delta x=\Delta y=$ 1 mm.}
    \label{time}
\end{figure}

As first analysis, we generate the time domain maps of the signal intensities, respectively chosen as maximum, minimum, and peak-to-peak, as shown in Figure \ref{time}. The best visualization in terms of contrast for S1 is given by the images reporting the minimum amplitude of the time waveforms (Figure \ref{time}b); while for S2 is given by the map of the maximum amplitude (Figure \ref{time}d).

In the third sample (S3), the image obtained from the intensity of the maximum and peak-to-peak analysis of the temporal waveforms suggests the presence of superimposed layers not visible to the naked eye (Figure \ref{time}g and \ref{time}i) while the map obtained from the minimum of the time signals highlights the presence of the cover layer (Figure \ref{time}h). The differences evidenced from the temporal waveform analysis can only suggest the presence of superimposed layers without really discriminating the hidden image of the base layer from the other two superimposed layers yielding, therefore a poor contrast image quality resulting in a difficult interpretation of the image.

For this reason, given the hints from the temporal analysis,  in order to study the internal structure of S3, we first investigated its stratigraphy using the sparse deconvolution algorithm \cite{Reyes2024} (described in Section \ref{sparsedeconvolution}) evidencing the presence of the three multiple layers. Then, we studied the mock-ups in frequency domain comparing the  spectral intensities of each pixel $I(\nu)_{ij}$ of the spectral intensity matrix $\mathbf{I_{signal}}$ to our database identifying all the materials present. Then we mapped each pigment according to its unique absorption fingerprint that generates the 2D chemical matrix $\mathbf{A_{S_{k}}^p}$ (with k = 1,2,3 for each sample and p = 1,2,..., n for each pigment; e.g., $\mathbf{A_{S_{1}}^2}$ corresponds to the chemical matrix of the second pigment present on sample S1.) of the three samples (Figures \ref{T}, \ref{H}, \ref{z_freq}). To do this, we developed an algorithm that verifies for a given pixel (${I(\nu)_{ij}}$), the resonance of a specific material, and stores $L_{signal}$ that is the area of $I(\nu)_{ij}$ in a specific frequency range that includes the absorption peak ($v_{0} \pm \Delta \nu $ ). The $L_{signal}$ value is stored in the corresponding matrix element ($A_{ij}$) of the final absorption matrixes ($\mathbf{A_{S_{k}}^p}$). To assess if the detection of a given material occurs, we chose to compare $L_{signal}$ to a threshold level ($Thr_{signal} = 3 \cdot L_{reference}$): we assume the presence of the peak only if its area value is at least three times greater than the one ($L_{reference}$) calculated from the reference spectrum ($I_{reference}$), collected on the support where no pigments are present, in the same frequency range. Selecting three times the area of the reference spectrum allows us to reduce the number of "false assessment". 
The value $v_{0} \pm \Delta \nu $ defining the integration range is chosen to maximise the contrast of the image: $v_{0}$ is chosen so that $||(I_{reference}(\nu_0)-I_{signal}(v_0)||$ is maximum which generally happens in correspondence to an absorption resonance centered at the frequency $\nu_0$, as explained in the Section \ref{dataanalysis}.

For S1 and S2 this procedure allows to separate the letter from the background, see Figures \ref{T} and \ref{H}; for S3 the procedure allows to reconstruct with a good image quality for each layer revealing the concealed text within the mock-up (i.e., letter "Z" in orpiment), see Figure \ref{z_freq}. 

In the following sections, we describe the results of the frequency analyzes for the three samples. 
Each two-dimensional map underwent visual inspection to identify pixels exhibiting potential algorithmic output errors (i.e., "false assessment"). This corrective procedure aimed to mitigate "overestimated" pixels, wherein the algorithm falsely identified pigment presence, and to rectify "underestimated" pixels, where the algorithm failed to detect the pigment's characteristic peak. 
In fact, we deliberately chose an algorithm that favours the inclusion of false positives over the omission of true signals. This conservative approach ensures that no meaningful signal is lost, particularly in boundary areas. The false positives introduced under this strategy are predominantly localized and do not significantly affect the interpretability of the final image image (see “Discussion” section).
For the correction of "underestimated" pixels, upon confirmation of the target pigment peak's presence, the intensity value was assigned as the mean value of the nearest neighbour pixel intensities. Conversely, for "overestimated" pixels, following verification of the peak's absence, the intensity value was set to zero.

\subsubsection{Samples S1 and S2}

\begin{figure}[h!]
    \centering
    \includegraphics[width=\linewidth]{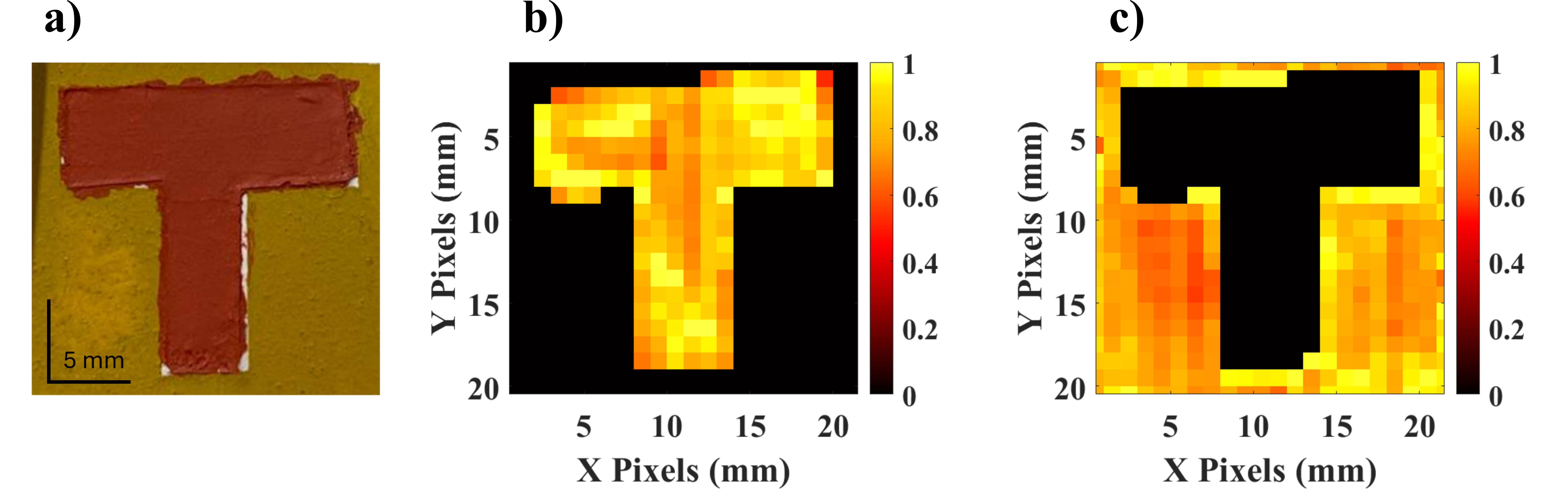}
    \caption{ a) Visible image of sample S1 letter "T"  in cinnabar and background of realgar; THz chemical mapping of thin layers of sample S1 obtained integrating: b) cinnabar peak centred at 1.13 THz and c) realgar peak centred at 1.48 THz. Each pixel dimension correspond to $\Delta x=\Delta y=$ 1 mm. The pixel values have been normalized for improved visualization.}
    \label{T}
\end{figure}

For sample S1 the broad peak centered at $\nu_C$=1.13 THz was selected to identify cinnabar (Figure \ref{T}b), since it offers the highest signal-to-noise ratio (SNR) and is better separated in the frequency spectrum from realgar fingerprints. The presence of cinnabar absorption was determined by integrating the absorbance spectrum within the range $\nu_C$ $\pm$ 0.02 THz. 
The obtained results allowed to clearly identify the red letter and selectively remove the background within an error referred to overestimated presence of cinnabar in 2.7\% of the total number of pixels which were corrected by visually checking each spectra (Fig. \ref{T}a). 
By following the same procedure, we can selectively identify realgar, in sample S1, selecting the peak with the higher SNR centered at $\nu_R$=1.48 $\pm$ 0.02 THz (Figure \ref{T}b). The background was selectively separated within a discrepancy referred to overestimated and underestimated presence of realgar in less than 1.8\% and 1.2 \% of the total number of pixels, respectively. These discrepancies are refined by visually inspecting each pixel's spectrum (Fig. \ref{T}c).

\begin{figure}[h!]
    \centering
    \includegraphics[width=\linewidth]{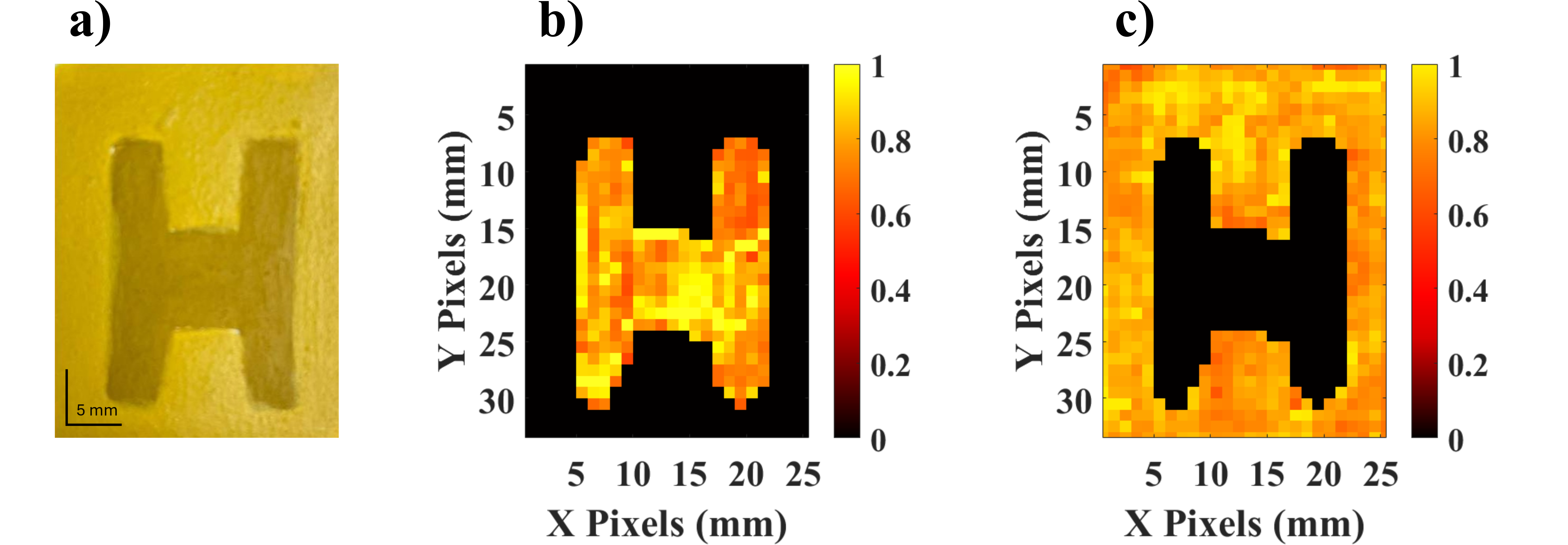}
    \caption{ a) Visible image of sample S2 letter "H" in realgar and background of orpiment; THz chemical mapping of thin layers of sample S2 obtained integrating: b) realgar peak centred at 1.48 THz and c) orpiment peak centred at 1.59 THz. Each pixel dimension correspond to $\Delta x=\Delta y=$ 1 mm. The pixel values have been normalized for improved visualization.}
    \label{H}
\end{figure}

In the second mock-up (S2), we investigated the ability of the proposed approach to selectively discriminate two arsenic sulfides (realgar and orpiment). 
The background constituted by orpiment was retrieved from its fingerprints centered at $\nu_O$=1.59 $\pm$ 0.02 THz (Figure \ref{H}b) while in the case of the "H" letter realised in realgar we employed the absorption peak at $\nu_R$=1.48 $\pm$ 0.02 THz (Figure \ref{H}b). 
The obtained results allowed to clearly identify the letter and selectively remove the background with a manual correction of the pixels within a discrepancy referred to overestimated presence of realgar in approximately 7.3\% of the total number of pixels (Figure \ref{H}a). The pigment constituting the letter "H" was identified with an underestimation of around 0.5\%.
Conversely, the background's pixels were overestimated of around 0.7 \% and underestimated of around 4.8\% of the total  (Fig. \ref{H}c); also in this case, before image correction, single spectra were visually checked and confronted with the orpiment's fingerprints.

\subsubsection{Sample S3}


In order to analyze the internal structure of S3, an investigation of its stratigraphy is included, as discussed below. In a multilayered sample, the THz reflected signal contains the superposition of several echoes corresponding to the internal interfaces.
The sparse deconvolution algorithm is applied to the raw data to identify these echoes \cite{Reyes2024}, therefore evidencing the thin layers separation that are undetectable in the original time profile. As consequences, the sparse deconvolution allows differentiating between attenuation levels resulting from multiple superimposed colour layers and those arising from fewer layers of increased thickness, thereby yielding critical information regarding the potential presence of concealed imagery. Subsequent spectral comparison of these points with our database of pure materials' spectral response facilitated the identification of specific pigments that contribute to the observed attenuations. The number of layers revealed by the deconvolution is consistent with the number of pigments identified by spectral analysis: cinnabar, orpiment, and realgar



\begin{figure}[h!]
    \centering
    \includegraphics[width=0.8\linewidth]{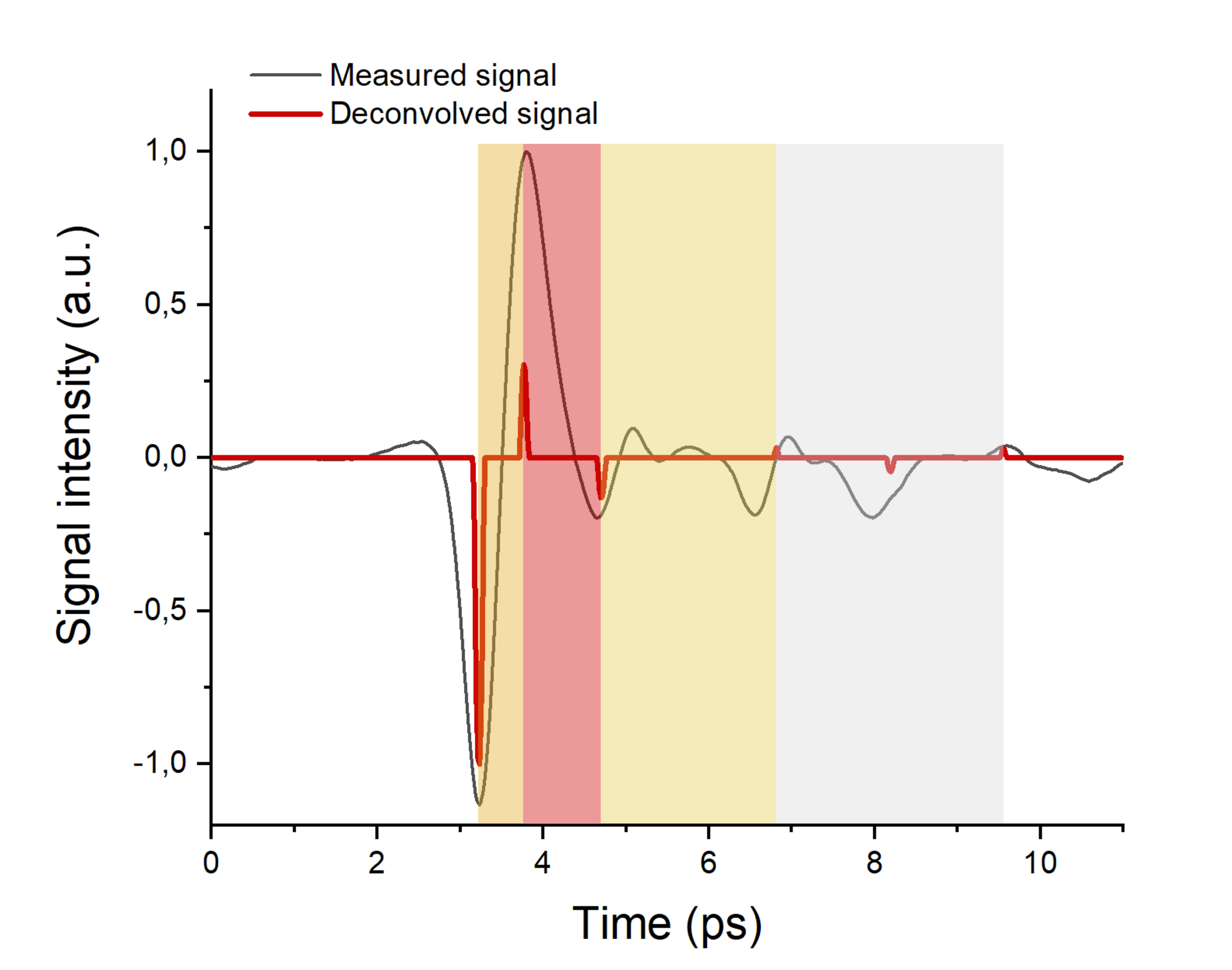}
    \caption{Raw signal and sparse deconvolved signals for one of the highly absorbing points of sample S3, originally collected within a 200 ps time window. This figure displays a zoomed-in 11 ps segment that highlights the portion containing information relevant to the stratigraphic layers, thereby emphasizing the temporal region of interest. The four layers of the mock-up sample are reported. The dark yellow region (\(\Delta\)t= 0.56 ps) correspond to the realgar layer (top layer, letter "C"), the red (\(\Delta\)t=0.94 ps) region correspond to the cinnabar layer (cover layer), the yellow region (\(\Delta\)t=2.1 ps) correspond to the orpiment layer (base layer, letter "Z"), and the grey region (\(\Delta\)t=2.75 ps) correspond to the support layer (paper). The presence of an echo in the paper layer is attributed to the possible presence of an internal defect that generates a reflection echo. Layer thicknesses were determined from sparse deconvolution echoes and the refractive indices of the materials ($n_{Orpiment}$=2.44, $n_{Cinnabar}$=1.87, $n_{Realgar}$= 1.92 for realgar, and $n_{Paper}$= 1.93.}
    \label{deconv}
\end{figure}

The sparse deconvolution algorithm identified five interfaces (Figure \ref{deconv}) corresponding to the three layers of pigments and the substrate (paper). The THz electric field presents a phase shift at each interface when the refractive index ($n$) of the subsequent layer is greater than the previous one. This corresponds to a change in the sign of the echo retrieved from the sparse deconvolution. The average values of the refractive indices, calculated from our experimental measurements in the THz range, are: $n_O$=2.44 for orpiment, $n_C$=1.87 for cinnabar, $n_R$=1.92 for realgar, and $n_{Paper}$=1.93 for the paper support. Taking into account the refractive indices, we retrieved the thickness of each pictorial layer: 52 $\pm$ 15 $\mu$m for the top layer (letter "C"), 55 $\pm$ 19 $\mu$m for the cover layer in cinnabar, 105 $\pm$ 24 $\mu$m for the base letter ("Z"), and 195 $\pm$ 17 $\mu$m for the support layer (paper). The observed presence of an echo within the paper layer (reported in the grey area in Figure \ref{deconv})  can be attributed to the potential existence of an internal defect \cite{Zhai2021}.

\begin{figure}[h!]
    \centering
    \includegraphics[width=0.95\linewidth]{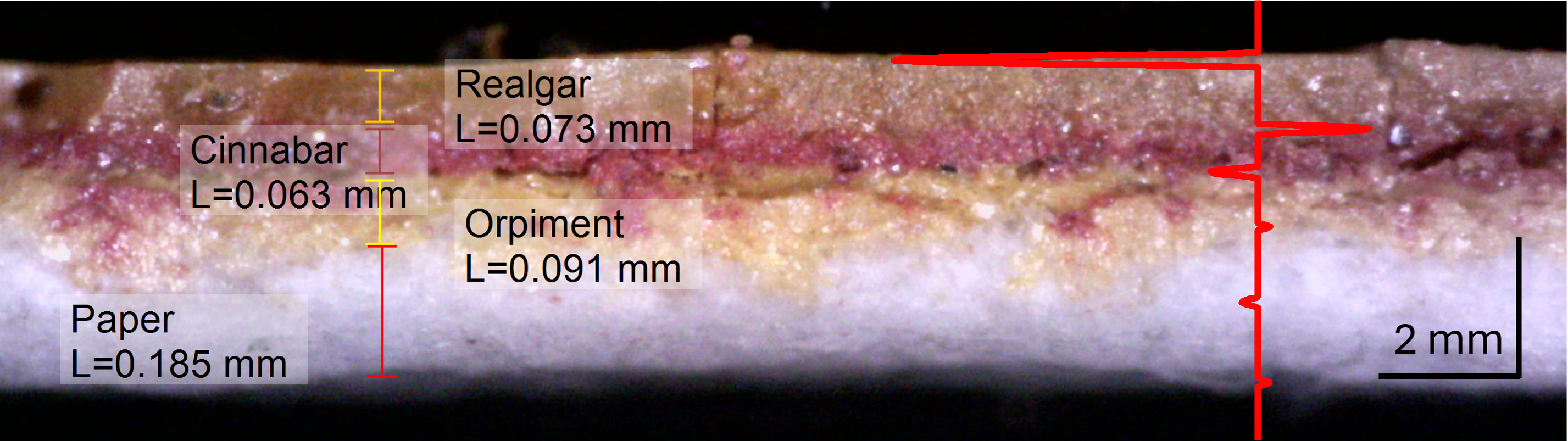}
    \caption{Microphotograph of a cross-sectional layering of sample S3 in which are reported the layers' thickness of: paper substrate (0.185 mm), base layer realized with orpiment (0.103 mm), cover layer in cinnabar (0.074 mm) and the top layer of realgar (0.093 mm) measured with the digital microscope.}
    \label{section}
\end{figure}

The cross-sectional image obtained with the digital microscope (Fig. \ref{section}) allowed us to confront the results obtained from the sparse deconvolution with the effective layers' thickness, showing a good agreement between the data (as also reported in Table \ref{tabsec}). The error in the thickness values retrieved from the sparse deconvolution represents the discrepancy in different points of measure due to the non-uniformity of the sample, as also visible in Figure \ref{section}.

\begin{table}[h!]
    \centering 
    \resizebox{\linewidth}{!}{
    \begin{tabular}{c|c|c|c|c}
         \textbf{Letter} & \textbf{Pigment/Support} &\textbf{Layer number} & \textbf{Average thickness $\pm$ $\sigma$} & \textbf{SD thickness $\pm$ $\sigma$}\\
         z & Orpiment & 1 &102 $\pm$ 24 $\mu$m& 105 $\pm$ 24 $\mu$m \\
         - & Cinnabar &2 & 60 $\pm$ 24 $\mu$m& 55 $\pm$ 19 $\mu$m \\
         C & Realgar & 3& 75 $\pm$ 30 $\mu$m& 52 $\pm$ 15 $\mu$m \\
         - & paper & support & 206 $\pm$ 17 $\mu$m & 195 $\pm$ 17 $\mu$m \\
         \end{tabular}
         }
    \caption{Comparison between the average value of the effective layers' thickness retrieved from the cross-sectional image obtained with the digital microscope and the average thickness values of each layer of sample S3 retrieved with the sparse deconvolution algorithm. The standard deviation values ($\pm\sigma$) obtained from measurements on cross-sectional images and from the sparse deconvolution algorithm are calculated at multiple points of sample S3. This value represents the heterogeneity of sample S3, where each layer exhibits non-uniform thickness.}
    \label{tabsec}
\end{table}



\begin{figure}[p]
    \centering
    \includegraphics[width=\linewidth, height=0.8\textheight, keepaspectratio]{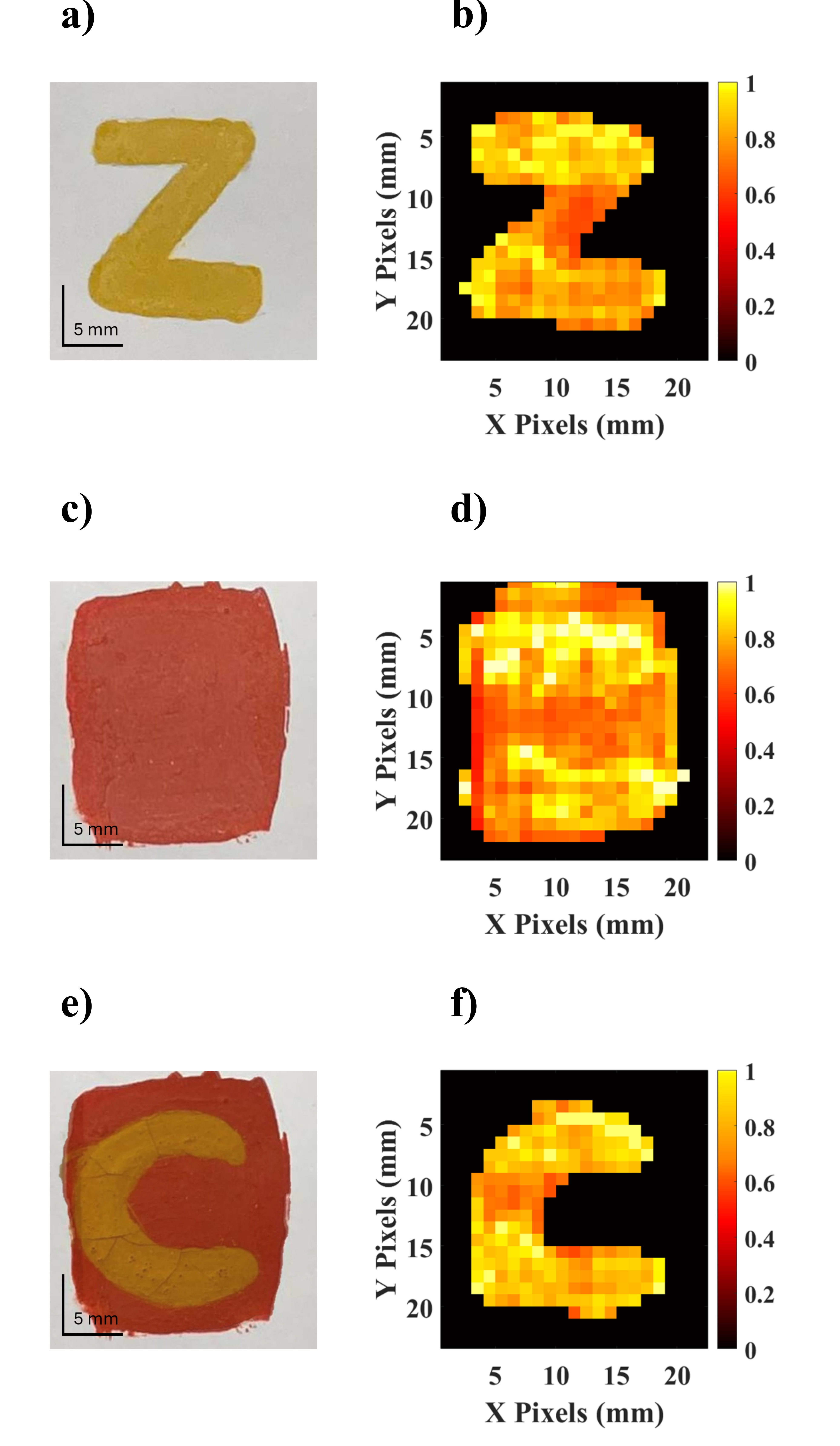}
    \caption{Sample constituted by superimposed layers (S3): a) visible image of base layer (letter "Z") in orpiment; b) THz chemical mapping obtained integrating orpiment peak centred at 1.59 THz; c) visible image of cover layer in cinnabar; d) THz chemical mapping obtained integrating cinnabar peak centred at 1.13 THz; e) visible image of top layer (letter "C") in realgar; f) THz chemical mapping obtained integrating realgar peak centred at 1.48 THz. Each pixel dimension correspond to $\Delta x=\Delta y=$ 1 mm. The pixel values have been normalized for improved visualization.}
    \label{z_freq}
\end{figure}

By employing the precise methodology detailed in the "Mock-ups mapping" section, we achieved the generation of three unequivocally distinct two-dimensional spatial maps. Each map rendered the artistic imagery corresponding to its specific stratum. As clearly demonstrated in Figure \ref{z_freq}, these three images exhibit marked differentiation and exceptional contrast, facilitating the unambiguous deciphering of the specimen's lettering, notably revealing internal layer details otherwise imperceptible to the unaided eye. This analytical approach yielded a demonstrably superior resolution compared to conventional time-domain analysis, a fact vividly illustrated by the comparative analysis of Figures \ref{time} and \ref{z_freq}.
The intensity in the map represents different absorption levels which evidence the non-uniformity of each layer, providing information on the material's distribution.

To retrieve the base layer of the sample (letter "Z" in orpiment) we used the algorithm selecting the peak centered at 1.59 THz. The visual inspection of the output from the algorithm gave a correction of 6.5\% for underestimated pixels and 11.6\% for overestimated pixels (Figure \ref{z_freq}b).
To recover the cover layer (cinnabar), we selected the peak centered at 1.13 THz. The pixels' distribution presents a disagreement around 1.6\% (underestimated) and 2.7\% (overestimated) of the total (Figure \ref{z_freq}d).
To retrieve the top layer (letter "C" in realgar), we used the algorithm selecting the peak centered at 1.48 THz; the discrepancy in reconstructing the realgar layer (letter "C") is approximately of 4.5\% (underestimated) and 12.7\% (overestimated) of the total (Figure \ref{z_freq}f).
These findings demonstrate that the methodology delineated herein facilitates the concurrent reconstruction of concealed textual elements at discrete depths within superimposed strata, alongside the spectroscopic identification of pigments employed within each respective layer, all derived from a singular analytical measurement. This approach thus presents a robust and non-invasive analytical methodology.

\section{Discussion}

The presented THz configuration and methodology effectively demonstrate their capability for the non-destructive investigation of pictorial materials. Specifically, THz multispectral imaging enabled the recovery of concealed textual information within multilayered pictorial structures. This reconstruction was achieved through frequency-domain analysis, leveraging the distinct spectral signatures of inorganic pigments and their subsequent identification via database correlation.

The emitted THz radiation exhibits a Gaussian spatial distribution with a $\sigma$ of approximately 500 $\mu$m at the focal plane. Consequently, the reconstruction of letter boundaries is subject to discrepancies of about 1 mm ($\Delta x=\Delta y= 1 mm$) in the retrieved letter dimensions. In fact, if a sampling measurement occurs by one spatial increment ($\Delta x$ or $\Delta y$) beyond the letter's edge, the substantial beam waist results in a detectable absorption signal due to the inherent spatial overlap between the beam spot and the first-neighbouring matrix pixels. To mitigate this effect, the setup can be implemented with optimized optics to reduce the beam focal size to approximately 100 $\mu$m. Furthermore, super-resolution techniques, such as convolutional neural networks and knife-edge (KE) methods, can be employed for imaging finer details \cite{Lei2021, Yuan2022, Vazquez2024, Wang2025}.

We performed two-dimensional mapping of three mock-ups by utilizing the spectral signatures of selected pigments. Our results demonstrate that images within thin-layered pictorial materials can be reconstructed, and non-uniform layer thicknesses can be retrieved through spectral band integration, providing crucial insights into material distribution.

To minimize discrepancies in reconstructed maps, a threshold level of
(\(\left(Thr_{signal} = 3 \cdot L_{reference}\right)\)) was adopted, optimizing pigment identification by reducing false positives caused by intensity variations in thicker absorbing layers. This approach ensures reliable pigment identification, with an average underestimation across all reconstructed images remaining below 5\%.

The observed discrepancy of approximately 13\% is primarily attributed to the presence of overestimated pixels (i.e., pixels where a signal is detected despite no actual pigment being present). This typically occurs near structural boundaries, such as the edges of letters, where the beam partially overlaps multiple materials.
This effect is closely tied to the beam waist size, which limits spatial resolution. In regions where adjacent materials meet, the THz beam can collect mixed signals, leading to false positives. As a result, reconstructed features, such as the letter shape, may appear approximately one pixel larger than their true dimensions in the mock-up.

The algorithm can be further refined by integrating chemometric approaches, such as principal component analysis, to preliminarily cluster pixels based on material composition, thereby mitigating discrepancies and enhancing classification accuracy.

Our multispectral mapping approach, driven by a custom-developed algorithm, provides precise reconstruction of hidden images in multilayered samples through frequency spectral analysis. This method significantly enhances stratigraphic analysis by simultaneously revealing concealed layers, determining their thickness, and selectively identifying their material composition. We validate its reliability using optical microscopy and sparse-deconvolution numerical analysis, demonstrating its effectiveness in both uncovering concealed images and accurately determining layer thickness.

This THz methodology offers a significant advancement over established diagnostic techniques. Notably, it overcomes limitations inherent in methods such as X-ray fluorescence (XRF), which encounters challenges in differentiating layers with similar elemental compositions (e.g., arsenic sulfides like orpiment and realgar), and Raman spectroscopy, which is often hampered by organic material interference \cite{DalFovo2021, Vandenabeele2007, Pause2021}. The THz approach, uniquely insensitive to binders and varnishes, provides accurate, single-measurement identification of concealed layers. This rapid, single-measurement capability translates to significantly shorter acquisition times compared to techniques
like spatially offset Raman spectroscopy (SORS) spectral mapping \cite{Vermeulen2025}, making it a highly efficient and effective tool for the non-destructive analysis of heritage materials. 
This study underscores the significant potential of in-situ THz multispectral imaging for revealing hidden details and advancing material analysis for historical preservation. The completely non-invasive nature of the proposed methodology, coupled with its insensitivity to high layer turbidity, establishes THz multispectral imaging as a powerful tool for unlocking concealed information within Cultural Heritage artifacts, expanding the frontiers of conservation science.

\section{Materials and methods}
\subsection{Terahertz time-domain spectroscopy (THz-TDS)}
\label{methods}
In this paper, a portable THz-TDS spectroscopic system, which operates in the 0.1-6 THz frequency range (approximately 3-200 $cm^{-1}$), was used. The system consists of an all-fiber-based femtosecond laser system. The laser pulses are emitted at a wavelength of 1560 nm and the repetition rate is 80 MHz. The pulses have a half-width of approximately 80 fs. The output beam is divided by a 50:50 fiber splitter in the emitter and detector branches. 
The optical pulses are converted into THz pulses through a photoconductive antenna, which serves as the emitter, the THz beam is focused through a system of four off-axis parabolic mirrors and the signal it is recollected by another photoconductive antenna acts as detector. 
The transmission setup, employed for the characterization of pure pigments, comprises a parabolic mirror (2" focal length) which collimates the emitted radiation, which is then focused onto the sample by the second mirror (4" focal length). The transmitted signal is recollimated by a third parabolic reflector (2" focal length) and subsequently focused onto the receiver by the fourth mirror (3" focal length), ensuring efficient signal transmission and detection.
In reflection configuration, the two mirrors next to the antennas have a focal length of 2" and the two central mirrors have a focal length of 4". They are slightly tilted, resulting in an incident angle of the THz beam on the sample of 8 degrees. In this configuration, the emitter and the detector along with all the optical components are enclosed in a reflection head. 
All the measurements were performed in ambient air since culturally valuable objects cannot always be subjected to purged conditions \cite{Koch2023}; thus, in this case, water vapour contributions are still present in the signal.
For its specific application to painting mock-ups, it was decided to work in reflection mode, by synchronizing a mechanical scanner (X-Y) with the THz acquisition, in order to acquire and map the signals. 
In this configuration, the reachable bandwidth approaches 3 THz with a number of averages equal to 100, chose to minimise the scanning time. Higher frequencies up to 5 THz can be reached by increasing the average value to 1000.

\begin{figure}[h!]
    \centering
    \includegraphics[width=0.5\linewidth]{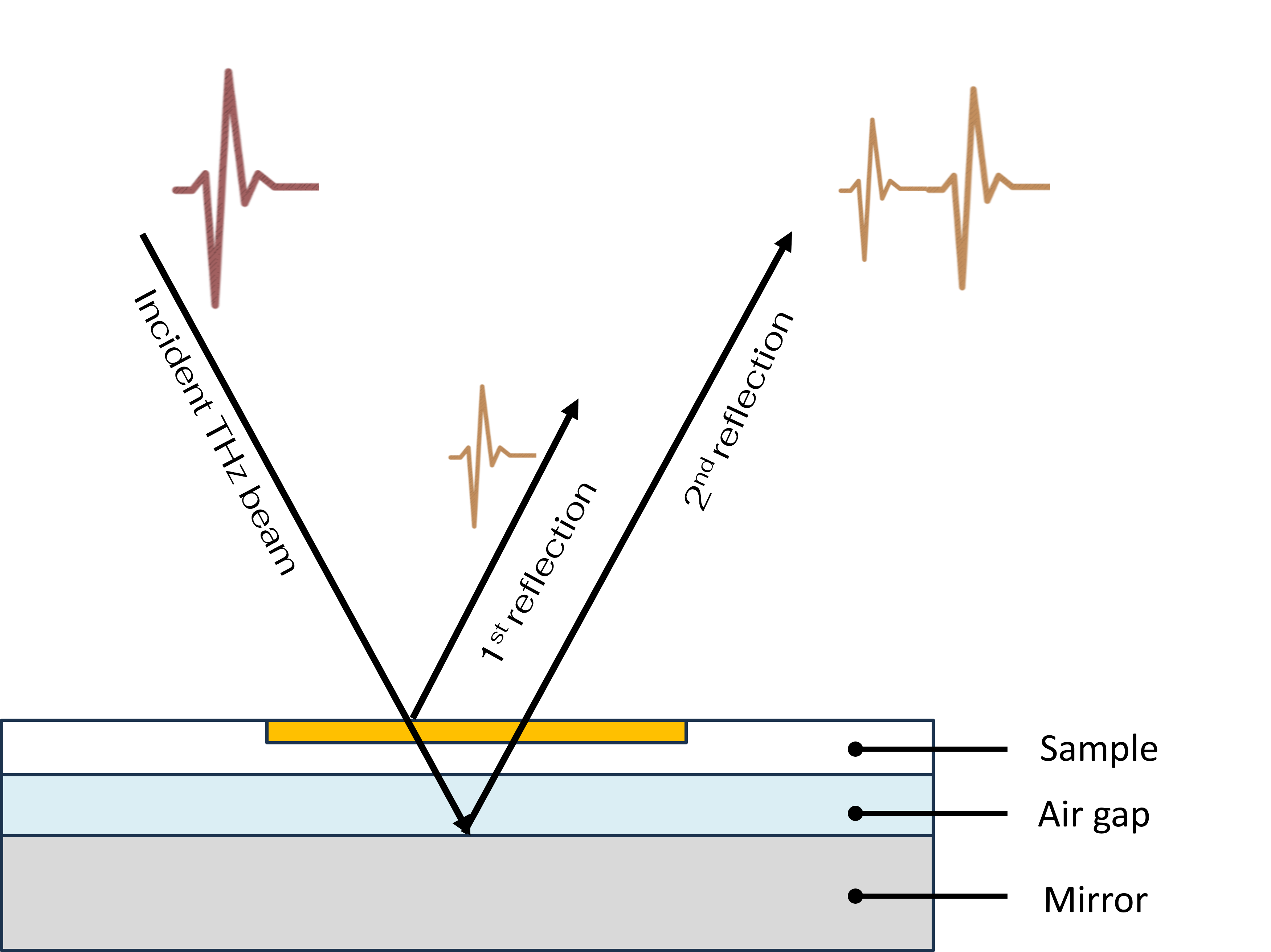}
    \caption{Schematic representation of the proposed set-up for the investigation of paper mock-ups through terahertz time-domain spectroscopy (THz-TDS).}
    \label{setup}
\end{figure}

The spot size of the THz beam was measured with MICROXCAM-384i, an uncooled microbolometric detector array composed of 384 x 288 pixel with each pixel constituted by a microbolometer (with a pitch of 35 $\mu$m).
The THz radiation emitted by the PCA has an approximately Gaussian profile. 
For the analysis, the samples were placed in the focal plane where the measured THz beam profile has $\sigma$=500 $\mu$m. 
According to the standard convention for Gaussian beams, the 1/$e^2$ beam diameter is given by 2w=4$\sigma$, which results in a beam diameter of approximately 2 mm. This value represents the effective spot size interacting with the sample. Given this 2 mm beam diameter, we selected a raster scan step size of 1 mm in both the x and y directions ($\Delta x$ = $\Delta y$= 1 mm), assuming isotropic resolution. This choice aligns with common guidelines for spatial sampling, which recommend a step size in the range of one-half to one-third of the beam diameter to ensure adequate spatial resolution. Specifically, to satisfy the Nyquist criterion, the sampling interval should be no larger than half the beam diameter.
Therefore, the mapping was performed with a spatial resolution of 1 mm for both $\Delta x$ and $\Delta y$.
The sample was placed at a fixed distance from a silver-coated mirror; consequently, the temporal waveform of the sample contains a sequence of signal echoes following the main reflected pulse (Fig. \ref{setup}).

\subsection{Data analysis}
\label{dataanalysis}

Chemical maps ($\mathbf{A_{S_{k}}^p}$, with k = 1,2,3 for each sample and p = 1,2,..., n for each pigment) have been reconstructed using an approach based on the integration of the spectral fingerprints of each material \cite{Prati2010, Rosi2011}. 
The THz electric field was recorded as a function of time and then the spectral response was obtained by Fourier transforming the filtered signal for each pixel ($I_{ij}(\nu)$). These frequency signals for each given point constitute the spectral intensity matrix $\mathbf{I_{signal}}$. Then, the custom developed algorithm is used to identify all the materials present in each pixel by comparing the spectral intensity $I(\nu)_{ij}$ to the homemade database.
Then, a frequency range that includes the absorption peak of interest ($\nu_0 \pm \Delta \nu$) is chosen to maximize the image contrast as:

\begin{equation}
   max||(I_{reference}(\nu_0)-I_{signal}(v_0)||
\end{equation}

Within this range, we assume the presence of the peak only if simultaneously the area value  $L_{s}$ is at least 3 times greater than the one obtained on a reference spectrum  $L_{r}$ collected on the support where no pigments are present: 

\begin{equation}
    L_{s,r} = \int_{\nu_i}^{\nu_f} I_{ij}(\nu)d\nu 
\end{equation}

where $s,r$ indicates the $L_{signal}$ and $L_{reference}$, $\nu_i$ and $\nu_f$ represent the minimum and the maximum frequency of interest, and $I_i (\nu)$ is the spectral intensity for each given pixel. 

\begin{equation}
    L_{signal} \geq Thr_{signal}= 3 \cdot L_{reference}
\end{equation}

If this condition is satisfied, the $L_{signal}$ value is stored in the corresponding matrix element (${A_{ij}}$) of the final absorption matrixes $\mathbf{A_{S_{k}}^p}$ for the corresponding sample.

\subsection{Sparse deconvolution}
\label{sparsedeconvolution}

The measured signal \( y(t) \) can be modelled as the convolution of the incident THz pulse \( h(t) \) with the system's impulse response \( f(t) \):

\begin{equation}
y(t) = h(t) \ast f(t) = \int_{-\infty}^{+\infty} h(\tau) f(t - \tau) d\tau
\end{equation}

In discrete form, this convolution is expressed as:

\begin{equation}
y_n = \sum_{m=0}^{M-1} h_m f_{n - m} + e_n
\end{equation}

where \( y_n = y(nT_s) \) is the measured signal, \( h_m = h(mT_s) \) represents the known incident pulse recorded from the reflection of the pulse on a metal mirror, \( f_{n - m} \) is the system's response to be recovered, and \( e_n \) accounts for noise. To recover the impulse response \( \mathbf{f} \) while promoting sparsity, we use Sparse Deconvolution with \( L_1 \) regularization, solving the following optimization problem based on \cite{Dong2017}:

\begin{equation}
\hat{\mathbf{f}} = \arg\min_{\mathbf{f}} \| \mathbf{H} \mathbf{f} - \mathbf{y} \|_2^2 + \lambda \| \mathbf{f} \|_1
\end{equation}

The first term ensures fidelity to the measured data by minimizing the \( L_2 \) norm of the residuals, while the second term, controlled by the regularization parameter \( \lambda \), promotes sparsity by applying the \( L_1 \) norm. This approach effectively reconstructs the system's impulse response \( f(t) \), even in the presence of noise, leading to a sparse and high-resolution representation of the signal.

\subsection*{Microphotography}
A portable digital microscope (AM7915MZT-Edge, Dino-lite, Almere, The Netherlands) was used to acquire images at 180–330× magnification.
The CMOS image sensor can transmit images at 15 fps with a resolution up to 1.3 Megapixel (1280 x 960).

\subsection{Pure pigments}
\label{pigments}
Cinnabar (\#10610), orpiment (\#10700) and realgar (\#10800) were purchased from Kremer Pigments Inc. (Munich, Germany). 
For their spectral characterisation (in transmission configuration) the samples were pressed into containment bolts in their pure powder form, without undergoing additional grinding, purification, or the incorporation of other materials (such as high-density polyethylene or microfine polytetrafluoroethylene). Multiple replicas were produced for each sample to evaluate data repeatability. Prior to each measurement, the pellets were carefully inspected with a digital microscope to ensure their structural integrity. Sample thickness was measured using a digital calliper with a resolution of ±0.01 mm. The inner diameter of each containment bolt averaged 8.35 mm, providing adequate space to accommodate the THz beam without causing clipping or edge diffraction.

\begin{table}[h!]
    \centering 
    \begin{tabular}{c|c|c|c|c|c}
        \textbf{Sample} & \textbf{Letter} & \textbf{Pigment} & \textbf{Background} & \textbf{Layering} &\textbf{Layer number} \\
         S1 & T & Cinnabar & Realgar & - & - \\
         S2 & H & Realgar & Orpiment & - & - \\
         S3 & Z & Orpiment & - & base & 1\\
         S3 & - & Cinnabar & - & cover &2 \\
         S3 & C & Realgar & - & top & 3\\
         \end{tabular}
    \caption{Information on mock-up samples' preparation}
    \label{tabsample}
\end{table}

\subsection{Mock-ups}
The same materials (described in Section \ref{pigments}) were then employed to produce mock-up samples on paper (Fabriano). The powder pigments were mixed with a binder (gum Arabic, Windsor \& Newton).
Schematic information on the mock-up samples' preparation are reported in Table \ref{tabsample}.

\begin{figure} [h!]
    \centering
    \includegraphics[width=0.5\linewidth]{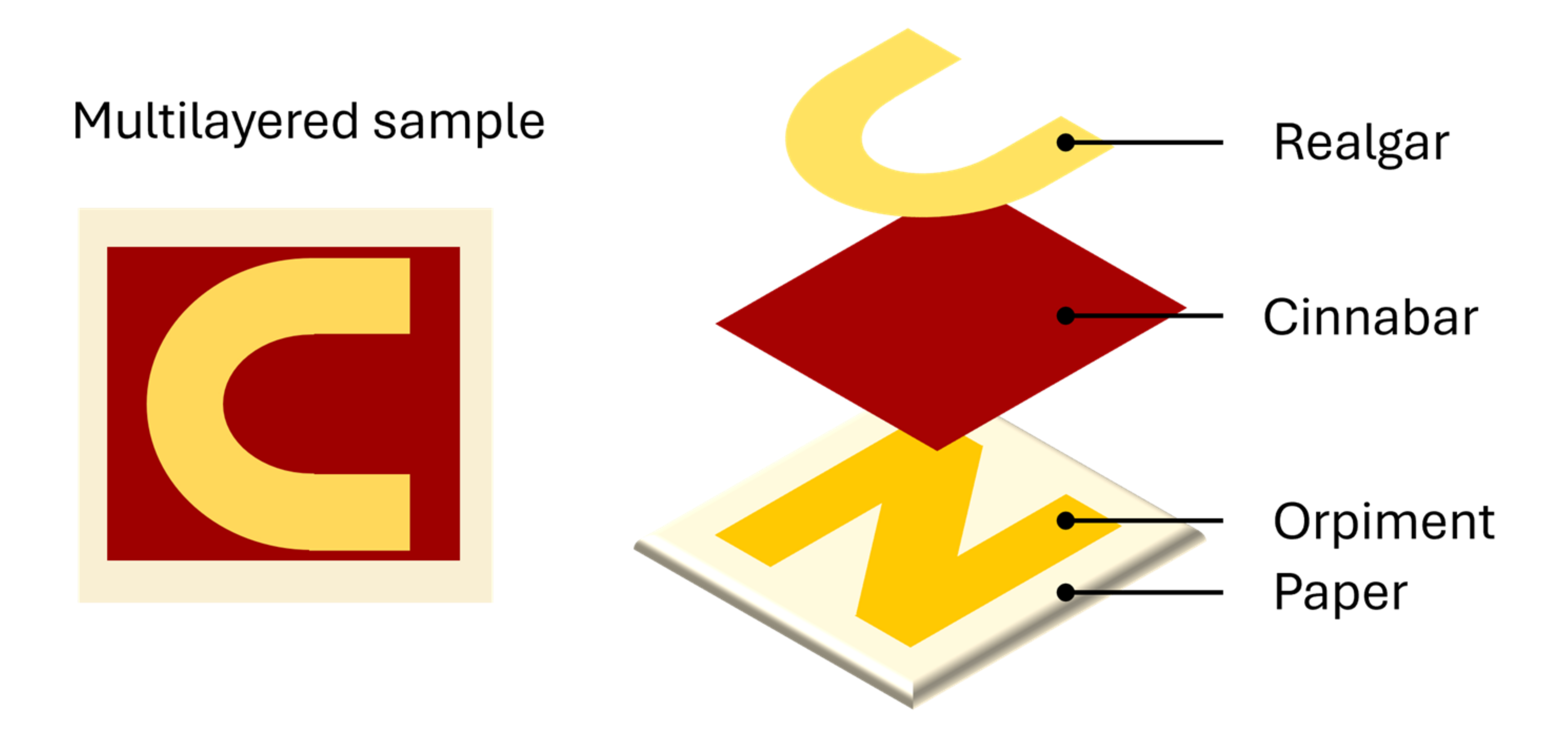}
    \caption{Schematic structure of the multilayered structure of samples S3 representing letter "Z" (base layer) in orpiment, cover layer in cinnabar, and letter "C" (top layer) in realgar.}
    \label{S3_schemesample}
\end{figure}

The first sample (S1) is constituted by a letter prepared with cinnabar ("T") surrounded by a yellow background of realgar (Figure \ref{T}a); thus, the sample was prepared with two chemically and spectral different materials to assess the validity of the proposed methodology. 
Sample S2 ("H") was realised with realgar and orpiment (Figure \ref{H}a) to prove that despite the chemical similarities, the two materials can be easily distinguished at THz frequencies. 

The multilayered sample (S3) was realised with three different pigments on paper (as reported in the schematic representation in Figure \ref{S3_schemesample}). Thus, we completely covered a sample prepared with orpiment (letter "Z", Figure \ref{z_freq}a) with cinnabar in order to make the letter not visible (Figure \ref{z_freq}c). Lastly, on the visible layer, we wrote a second letter ("C") with realgar (Figure \ref{z_freq}e).

\bibliography{Mapping}

\section*{Acknowledgements}

This work was carried out thanks to Sapienza Large Projects Research Call 2023 titled "TforCH: R\&D on the potentiality of THz radiation for Cultural Heritage". This work was conducted in the framework of "PRIN 2022: TREX a prototype of a portable and remotely controlled platform based on THz technology to measure the one health vision: environment, food, plant health, security, human and animal health" funded by the European Union - Next Generation EU (CUP B53D23013610006 - Project Code 2022B3MLXB PNRR M4.C2.1.1). This work was supported by Sapienza competitive grants: Grandi Attrezzature Scientifiche (2018) titled "SapienzaTerahertz: THz spectroscopic image system for basic and applied sciences". This research was also supported by "STORM - Sensori su sistemi mobili e remoti al Terahertz PNRM a2017.153 STORM" funded by Ministero della Difesa and "R-SET: Remote sensing for the environment by THz radiation" Large Research Projects of Sapienza, University of Rome.
The authors would like to thank "CSN5 Grants INFN Roma 1" for their contribution to this work.

\section*{Author contributions statement}

Conceptualization, C.M. and M.P.; methodology, C.M., D.F., A.C. and M.P.; software, C.M., D.F., A.C., M.B., L.P., M.M., and M.P.; validation, C.M., D.F., A.C., and M.P.; formal analysis, C.M., D.F. and M.P.; investigation, C.M., D.F. and M.P.; resources, C.M., and M.P.; data curation, C.M., D.F., and M.P.; writing---original draft preparation, C.M., and M.P.; writing---review and editing, all the authors; visualization, C.M.; supervision, C.M., and M.P.; project administration, C.M., and M.P.; funding acquisition, C.M., and M.P. \\
All authors have read and agreed to the published version of the manuscript.

\section*{Additional information}
\noindent
\textbf{Competing interests}: The authors declare no competing interests. 
\\
\textbf{Data availability}: The datasets used and/or analysed during the current study are available from the corresponding author on reasonable request.

\end{document}